\documentclass[aps,prx,twocolumn,showpacs,longbibliography]{revtex4-1}
\usepackage{amsmath}
\usepackage{amsthm, amssymb} 
\usepackage{bbold}                       
\usepackage{graphicx}
\usepackage{color}
\usepackage{times, verbatim}      
\usepackage{natbib}
\usepackage{hyperref}
\usepackage{accents}
\hypersetup{
       colorlinks=true,
}

\newcommand*{\dt}[1]{%
  \accentset{\mbox{\large\bfseries .}}{#1}}

\newcommand{\rholl}{\rho_{l^{\prime}l}^{\sigma}}
\newcommand{\dll}{\Gamma_{l^{\prime}l}^{\sigma}}
\newcommand{\gammall}{\gamma_{l^{\prime}l}^{\sigma}}
\newcommand{\rhokkp}{\rho_{\boldsymbol{k},\boldsymbol{k}'}^{\sigma}}
\newcommand{\gamkkp}{\gamma_{\boldsymbol{k},\boldsymbol{k}'}^{\sigma}}
\newcommand{\gamkpkd}{\gamma_{\boldsymbol{k}',\boldsymbol{k}}^{\sigma \dagger}}

\begin{document}

\title{Transport of Hubbard-Band Quasiparticles in Disordered Optical Lattices}

\author{ V. W. Scarola$^{1}$ and B. DeMarco$^{2}$}
\affiliation{$^{1}$Department of Physics, Virginia Tech, Blacksburg, Virginia 24061, USA}
\affiliation{$^{2}$Department of Physics, University of Illinois at Urbana-Champaign, Urbana, Illinois 61801, USA}

\begin{abstract}
Quantum degenerate gases trapped in optical lattices are ideal testbeds for fundamental physics because these systems are tunable,  well characterized, and isolated from the environment.  Controlled disorder can be introduced to explore suppression of quantum diffusion in the absence of conventional dephasing mechanisms such as phonons, which are unavoidable in experiments on electronic solids.  Recent experiments use transport of degenerate Fermi gases in optical lattices (Kondov et al. Phys. Rev. Lett. 114, 083002 (2015)) to probe a particularly extreme regime of strong interaction in what can be modeled as an Anderson-Hubbard model.  These experiments find evidence for an intriguing insulating phase where quantum diffusion is completely suppressed by strong disorder. Quantitative interpretation of these experiments remains an open problem that requires inclusion of non-zero entropy, strong interaction, and trapping.  We argue that the suppression of transport can be thought of as localization of Hubbard-band quasiparticles.  We construct a theory of transport of Hubbard-band quasiparticles tailored to trapped optical lattice experiments.  We compare the theory directly with center-of-mass transport experiments of Kondov et al. with no fitting parameters.  The close agreement between theory and experiments shows that the suppression of transport is only partly due to finite entropy effects.   We argue that the complete suppression of transport is consistent with Anderson localization of Hubbard-band quasiparticles.  The combination of our theoretical framework and optical lattice experiments offers an important platform for studying localization in isolated many-body quantum systems.
\end{abstract}

\date{\today}


\maketitle

\section{Introduction}

Understanding the motion of a quantum particle in an otherwise isolated lattice under the influence of an applied field is central to our understanding of conductivity in electronic solids.  The theory of Anderson localization \cite{anderson:1958,abrahams:2010} predicts that quantum diffusion of a single particle can fail in a disordered lattice.  Above a critical disorder strength, for which the mobility edge encompasses all states participating in transport \cite{mott:1974,mott:1987}, strong interference forbids quantum diffusion.  Anderson's mechanism of localization was first discussed in the context of as a simplified model designed to treat the propagation of highly excited states of nuclear spin systems but is has since been applied to a wide variety of other systems \cite{abrahams:2010}, including quantum degenerate atomic gases \cite{billy:2008a,roati:2008,chabe:2008,lemarie:2009,kondov:2011,jendrzejewski:2012}.  Disorder-induced localization is also believed to play a key role in metal-insulator transitions in a wide-range of materials \cite{mott:1974,mott:1987,abrahams:2010}.

\begin{figure}[t]
\includegraphics[clip,width=70mm]{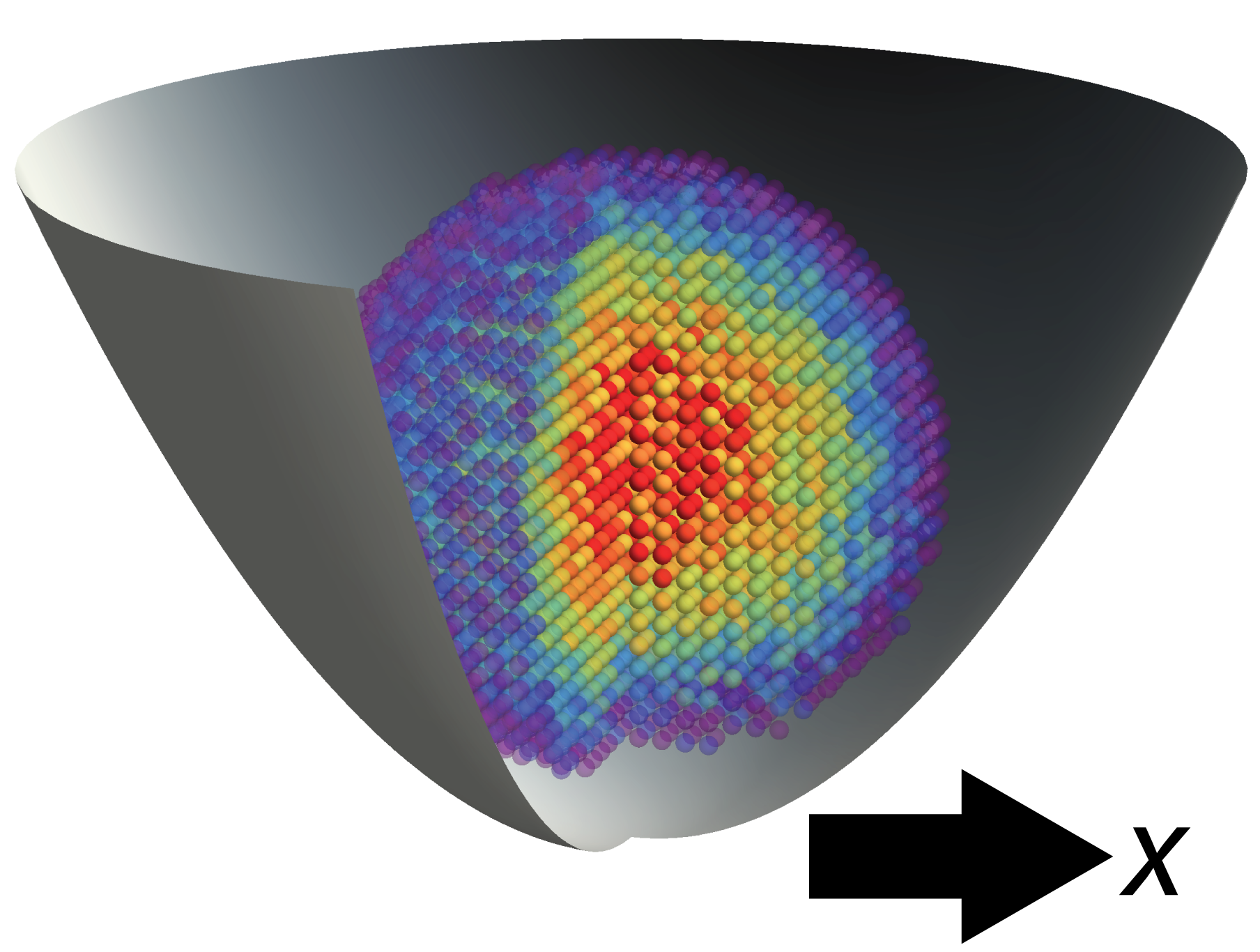}
\caption{(Color online)  Schematic showing disordered lattice sites in a parabolic trapping potential.  The site coloring represents a dense core that gives way to zero density at the edges.  The system studied here can be thought of as a strongly interacting high temperature paramagnet with a density less than one at the center.  An applied shift of the external trapping potential along the $x$-direction for a time $\tau=\tau_{P}$ forces center-of-mass motion along the $x$ direction only if the atoms are mobile.  $\tau_{P}$ is chosen to be short on the time scale of the inverse trapping frequency.}
\label{fig_lattice_schematic}
\end{figure}

Subsequent theoretical studies of Anderson localization found that inclusion of realistic effects, specifically inter-particle interactions and non-zero temperature \cite{finkelshtein:1983,lee:1985, vojta:1998,byczuk:2005,evers:2008,abrahams:2010}, pose prominent problems.  The competition between Anderson localization and strong interaction effects have been studied with a variety of methods, e.g., quantum Monte Carlo \cite{denteneer:1999}, dynamical mean field theory \cite{georges:1996,byczuk:2005,semmler:2010}, and related quantum cluster methods \cite{maier:2005}.  Refs.~\cite{byczuk:2005} and ~\cite{semmler:2010}, for example, found a correlated Anderson insulator ground state for large disorder strengths indicating that Anderson localization persists in a strongly interacting limit.  A more complete understanding of the interplay of strong inter-particle interactions and disorder is urgently needed to enhance our knowledge of strongly correlated materials such as high-temperature superconductors.

Related work by Basko et al. \cite{basko:2006} has triggered considerable interest in the interplay between interactions, temperature, and Anderson localization.  Their work indicates that a correlated Anderson insulator is stable at non-zero temperatures and corresponds to a many-body localized state.  This is surprising because one might expect that interactions lead to dephasing effects that mimic the effects of heat and particle number reservoirs \cite{moix:2013a} that are known to lead to conduction via variable range hopping in certain solids \cite{mott:1987}.  Interactions would be expected to lead to effective reservoirs even in the absence of an explicit reservoir.  But Ref.~\cite{basko:2006} argues, surprisingly, that interactions allow a correlated Anderson insulator to survive up to a characteristic temperature.  

Quantum degenerate gases of atoms trapped in optical lattices offer a controlled arena to study the interplay of interactions, disorder, and thermal effects \cite{jaksch:1998,greiner:2002,bloch:2008,sanchez-palencia:2010} because they are, to an excellent approximation, isolated.  The entropy per particle, controlled via cooling in a parabolic trap, determines an equilibrium temperature when the lattice is turned on, since atomic gases thermalize through inter-particle interactions \cite{rigol:2008,mckay:2011}.  As a result of their isolation, quantum degenerate Fermi gases in optical lattices exhibit quantum diffusion (see, e.g., Ref~\cite{schneider:2012}), even though their temperatures are a significant fraction of the Fermi energy.  This offers a useful regime to study because isolated systems can, in principle, exhibit many-body localization even at high temperature \cite{oganesyan:2007}.  Furthermore, optical lattice experiments are well characterized \cite{bloch:2008,zhou:2010}: interaction strength, lattice depth, entropy, density, and other parameters are all known and tunable.  The impact of disorder can therefore be studied independently of conventional dephasing phenomena arising from contact with reservoirs \cite{chen:2008,white:2009,pasienski:2010,gadway:2011,beeler:2012,brantut:2012,tanzi:2013,krinner:2013}.

Recent experimental work \cite{kondov:2015} has investigated interacting fermions confined in a cubic optical lattice to study the influence of quenched speckle disorder on center-of-mass transport, Fig.~\ref{fig_lattice_schematic}.  This system is accurately described using the Anderson-Hubbard model.  They find intriguing insulating behavior above a critical disorder strength that agrees qualitatively with the predictions for many-body localization in the weakly interacting regime \cite{basko:2006}.  Control over the lattice potential depth and disorder strength was shown to lead to a regime where known types of insulating behavior can be excluded.  For example, it was demonstrated that the insulating regime occurs for disorder strengths well below the classical percolation threshold \cite{kondov:2015}.  Furthermore, the system was made dilute enough to avoid forming Mott \cite{jordens:2008,schneider:2008,jordens:2010} and band insulators.  The regime they explored can be thought of as a strongly interacting Hubbard paramagnet with a temperature well below the bandwidth.  The insulating behavior of the isolated-strongly interacting particles in these experiments \cite{kondov:2015} is therefore a highly non-trivial probe of localization in many-body quantum states.

 In this article, we provide, to our knowledge, a new perspective on disorder-induced localization in the Anderson-Hubbard model and the measurements in Ref.~\cite{kondov:2015}.  We establish a connection to Hubbard-band quasiparticles using a direct comparison between theory and experiment with no fitting parameters.  This approach enables us to treat the strongly interacting limit, which was not possible using the perturbation theory employed in Ref.~\cite{basko:2006}. We derive the equations of motion for the Anderson-Hubbard model in the paramagnetic regime while taking into account all important experimental aspects, particularly trapping and finite entropy effects.  We show that the equations of motion derived here to include a trap reduce to the Hubbard-I \cite{hubbard:1963} approximation normally considered in the translationally invariant limit.  This demonstrates that our equations of motion quantitatively capture the dynamics of Hubbard-band quasiparticles in a trap. The Hubbard-I approximation is a strong coupling approximation that becomes exact in the limit of strong interactions and high entropy.  Our formalism is therefore directly applicable to the experiments of Ref.~\cite{kondov:2015}.   

We use parameters taken from Ref.~\cite{kondov:2015} to effectively replicate the experiment numerically and find evidence for a quasiparticle mobility edge.  We find that at low disorder strengths the quasiparticles propagate in the lattice under an applied force, i.e., they have non-zero mobility.  We also study the result of increasing disorder.   At large disorder strengths we identify a transition to an insulator through the absence of center-of-mass motion.  A direct comparison between theory and experiment shows good agreement.  We argue that the insulating behavior observed in Ref.~\cite{kondov:2015} is consistent with Anderson localization of Hubbard-band quasiparticles.  To our knowledge, disorder-induced localization in the Hubbard model has not been previously understood using this approach, which is complementary to other methods,  e.g., perturbative theory \cite{basko:2006} and dynamical mean field theory \cite{georges:1996,byczuk:2005,semmler:2010}.
 
We begin in Sec.~\ref{sec_model} by defining the model used to simulate the experiments of Ref.~\cite{kondov:2015} and all necessary parameters.  Here we also define the center-of-mass velocity as the key observable.  In Sec.~\ref{sec_dynamics} we then derive the equations of motion in the paramagnetic regime.  Sec.~\ref{sec_hubbardapp} then shows that the equations of motion reduce to the Hubbard-I approximation \cite{hubbard:1963} that was originally used to define Hubbard-band quasiparticles.  Here we also show that, in a strong coupling limit, Hubbard-band quasiparticles obey an effective Anderson model of non-interacting quasiparticles.  Sec.~\ref{sec_initial_state} then defines the approximations used in constructing the initial state that is propagated using the equations of motion.  Sec.~\ref{sec_heating} points out an important feature of the initial states used in these experiments.  We find that increasing disorder at fixed entropy effectively raises the system temperature to at most $B/3$, where $B$ is the bandwidth.   Even though this heating keeps the temperature well below the bandwidth, it is nonetheless an important aspect of these experiments that must be included to make a quantitative comparison with theory.  Sec.~\ref{sec_com_dynamics} presents our central results.  Here we directly compare numerical solutions of the equations of motion with experiments.  We find that low disorder allows the Hubbard-band quasiparticles to propagate in the trap.  But we find a critical disorder strength above which all transport is suppressed.  We conclude in Sec.~\ref{sec_discussion} by interpreting the results presented here as evidence for the Anderson localization of Hubbard-band quasiparticles.

\section{Model and Parameter Regimes}
\label{sec_model}
We study the dynamics of an equal population of two-component fermions in a cubic optical lattice in the presence of spatial disorder.  For deep lattices we assume that all $N$ particles reside in the lowest Bloch band.  In this limit the single-band Anderson-Hubbard model is an excellent approximation \cite{bloch:2008,kondov:2015}:
\begin{eqnarray}
\hspace{-0.2cm}
    H_{\text{AH}}=  \sum_{j,j^{\prime},\sigma} T_{j,j^{\prime}} c_{j,\sigma}^\dagger
    c^{\phantom{\dagger}}_{j^{\prime},\sigma}
    + U \sum _{j} n_{j,\uparrow} n_{j,\downarrow} +\sum_{j} \mu _{j} n_{j}
    \label{eq_hubbardmodel}
   \end{eqnarray}
Here $c_{j,\sigma}^\dagger$ creates a fermion of spin $\sigma\in \uparrow,\downarrow$ at a site $\mathbf{R}_{j}$, $U$ is a repulsive interaction, and 
$n_{j}=n_{j,\uparrow}+n_{j,\downarrow}$ is the number operator.  The matrix elements $T_{j,j'}\equiv -t\delta_{\mathbf{R}_{j},\mathbf{R}_{j'}+\boldsymbol{\delta}}$ are written in terms of a Kronecker delta, $\delta$, that enforces a hopping energy $t$ between nearest neighbor sites ($\boldsymbol{\delta}$ is a nearest neighbor bond vector.)

The last term in Eq.~(\ref{eq_hubbardmodel}) includes spatially inhomogeneous perturbations to the chemical potential.  We define:
\begin{eqnarray}
\mu _{j}=-\mu_{0}+\frac{m\omega^{2} a^{2} R_{j}^{2}}{2}+\epsilon_{j}+V_{P}\mathbf{R}_{j}\cdot \hat{x},
\label{eq_muj}
 \end{eqnarray}
where $\mu_{0}$ is the average chemical potential, $m$ is the atomic mass, $\omega$ is the trapping frequency that parameterizes the external confinement, 
$a$ is the lattice spacing determined by the optical lattice laser wavelength, $\epsilon_{j}$ denotes spatially random disorder, and $V_{P}$ is a pulse strength that is switched on for a time $\tau=\tau_{P}$ to effectively shift the trap center.   

$V_{P}$ acts as the analogue of a weak electric field used to drive transport along the $x$-direction, see Fig.~\ref{fig_lattice_schematic}.  At long times a single particle with no disorder will oscillate in the trap.  But we consider pulse times that are short with respect to the inverse trapping frequency to focus on the linear response regime probed in Ref.~\cite{kondov:2015}.  At these short times, the center-of-mass velocity is unidirectional and, in the absence of disorder, increases linearly with $V_{P}$.

We consider two distinct distributions of site disorder.  In the experiments of Ref.~\cite{kondov:2015} the speckle potential used to establish a disordered optical lattice creates an exponential probability distribution function for the on-site energies \cite{zhou:2010}:
 \begin{eqnarray}
 P_{E}(\epsilon)=\frac{e^{-\epsilon/\Delta_{E}}}{\Delta_{E}},
 \label{eq_PE}
  \end{eqnarray}
where $\Delta_{E}$ is the strength of the exponentially distributed disorder assuming $\epsilon>0$ (this is accurate to within 10\% of the disorder strength used in Ref.~\cite{kondov:2015}).  We also consider a uniform (boxed) disorder probability distribution function for the on-site energies $\epsilon_{j}$:
 \begin{eqnarray}
  P_{U}(\epsilon)=\frac{\Theta(\Delta_{U}/2-\vert\epsilon\vert)}{\Delta_{U}},
   \label{eq_PU}
   \end{eqnarray}
where $\Theta$ is the Heaviside step function and $\Delta_{U}$ parameterizes the strength of the uniformly distributed disorder.  $P_{E}$ introduces behavior that is distinct from more common models with $P_{U}$ because changing $\Delta_{E}$ at fixed $N$ forces $\mu_{0}$ to change.  This is in contrast to changes in $\Delta_{U}$ which leaves $\mu_{0}$ constant at fixed $N$.

Eq.~(\ref{eq_hubbardmodel}) quantitatively captures the essential properties of the experiments in Ref.~\cite{kondov:2015}.  We ignore disorder in $t$ and $U$ that was shown \cite{zhou:2010} to be narrowly Lorentzian distributed.  In what follows, we find that we are able to make quantitative comparison with experiment even while excluding the disorder in $t$ and $U$.  We will return to this point in Sec.~\ref{sec_hubbardapp}.

The experiments proceed by trapping a fixed number of particles with a fixed entropy, $S$.  The entropy and all other necessary model parameters were determined in Ref.~\cite{kondov:2015} and are shown in Table~\ref{tab1}.  We focus on the two lattice depths with high $U$, where $U/t\approx6$ and $U/t\approx9$ for $6 E_{R}$ and $7E_{R}$, respectively, which allows for a strong coupling approximation that becomes exact in the limit $U/t\rightarrow\infty$.  Table~\ref{tab1} leaves no fitting parameters in using approximate solutions of Eq.~(\ref{eq_hubbardmodel}) to compare with the experiments of Ref.~\cite{kondov:2015}.

\begin{table}[t]
\caption{Table of parameters used in the experiments of Ref.~\cite{kondov:2015}.  Here the recoil energy is $E_{R}=h^{2}/(8ma^{2})$ and the atomic species is $^{40}$K. } 
\centering
\vspace{0.1cm}
\begin{tabular}{|l|c|c|c|c|c|}
  \hline
  Lattice Depth           & $V_{L}$  		& $6 E_{R}$ & $7 E_{R}$          \\[-.1ex] \hline   \hline
   Trap Frequency		& $\omega$       & 	$110\times2\pi$Hz & 	$114\times2\pi$Hz  \\[-.1ex]\hline
    Lattice Spacing		& $a$      & 	$391.1$ nm & 	$391.1$ nm  \\[-.1ex]\hline
  Number of Particles & $N$		& 	$47100 \pm 6500$ & 	$48700 \pm 1900$  \\[-.1ex]\hline
 Entropy per Particle &  $ S/N$   	& 	$1.51 \pm 0.18 k_{B}$& 	$1.6 \pm 0.17 k_{B}$ \\[-.1ex]\hline
 Hopping 			& $t$     		& 	0.0509 $E_{R}$ & 	0.0395 $E_{R}$  \\[-.1ex]\hline
 Interaction 		& $U$    		 & 0.304 $E_{R}$ & 0.355 $E_{R}$  \\[-.1ex]\hline
Relative Strength		& $U/t$    		 & 5.97 & 8.98  \\[-.1ex]\hline
  Disorder 	Strength	& $\Delta_{E}$ 	& 	0-2 $E_{R}$ 	& 	0-2 $E_{R}$  \\[-.1ex]\hline
  Pulse Time 		& $\tau_{P}$ 	& 	2 ms  & 	2 ms  \\[-.1ex]\hline
  Pulse Strength		& $V_{P}$ 	& 	0.011 $E_{R}$ & 	0.011 $E_{R}$  \\[-.1ex]\hline
\end{tabular}
\label{tab1}
\end{table}

We will show that the entropies reported in Table~\ref{tab1} imply temperatures that are well above the N\'eel temperature, $\sim t^{2}/U$ \cite{staudt:2000,jordens:2010,fuchs:2011,paiva:2011,kozik:2013,hart:2015}.  The experimentally relevant temperature regimes are above the hopping but below the bandwidth.  Our central set of approximations in studying Eq.~(\ref{eq_hubbardmodel}) can be summarized by:
\begin{eqnarray}
 t&\ll& U  \nonumber \\ 
 t\lesssim &k_{B}T& < 12 t 
 \label{eq_approx}
 \end{eqnarray}
where the first inequality assumes that we focus on the high lattice depth data of Ref.~\cite{kondov:2015}, and the second inequality implies that high temperature limits are valid approximations.  Sec.~\ref{sec_initial_state} shows that the initial state for the parameters defined by Table~\ref{tab1} can be thought of as a dilute ($\langle n \rangle <1$) high temperature paramagnet.  We will therefore focus our study to strongly interacting paramagnetic regimes.

To make contact with experimental results presented in Ref.~\cite{kondov:2015} we study the dynamics of the center of mass.   The time dependent center-of-mass velocity in particular:
\begin{eqnarray}
\mathbf{V}_{\text{C.O.M.}}(\tau)=  \sum_{j} \mathbf{R}_{j} \langle \dt{\langle  n_{j}  \rangle}  \rangle_{D}  
\label{eq_vcom}
 \end{eqnarray}
 was inferred from time of flight images \cite{kondov:2015}.  Here $ \langle \langle ... \rangle  \rangle_{D} $ indicates disorder averaging of expectation values and $\tau$ denotes time.  In the following we find that disorder averaging over 25-50 realizations is sufficient to reach convergence in our numerical simulations.  We will use Eq.~(\ref{eq_vcom}) to compute the center-of-mass velocity along the direction of the applied pulse after a time $\tau=\tau_{P}$:
 \begin{eqnarray}
V_{\text{C.O.M.}}=\hat{x}\cdot\mathbf{V}_{\text{C.O.M.}}(\tau_{P}).  
\label{eq_vcom_scalar}
 \end{eqnarray}
This quantity is akin to measures of mobility in solids.  For example, in the Drude model of electrical conductivity, $V_{\text{C.O.M.}}$ is proportional to the electron mobility when measured after a pulse.   $V_{\text{C.O.M.}}$ will therefore offer a useful probe to study the impact of disorder on transport of strongly interacting atoms in optical lattices. 

\section{Dynamics from Equations of Motion}
\label{sec_dynamics}

To study the center-of-mass dynamics we derive equations of motion for correlation functions related to observables.  The trapping potential in Eq.~(\ref{eq_hubbardmodel}) breaks translational invariance.  We will derive the equations of motion in the site (Wannier) basis as opposed to the more conventional $k$-space (Bloch) basis to allow studies of local dynamics in trapped lattices.  We approximate the equations of motion by relying on the strong interaction/high temperature limit, Eqs.~(\ref{eq_approx}).  The next section shows that our approximation reduces to Hubbard's decoupling of the equations of motion, the Hubbard-I approximation \cite{hubbard:1963}, that introduced the concept of Hubbard-band quasiparticles.  The equations of motion derived here therefore offer a tool to study the local dynamics of Hubbard-band quasiparticles in the absence of translational invariance.  

The exact equations of motion for the charge and spin degrees of freedom are given by:
\begin{eqnarray}
i\hbar \frac{d \langle\rholl \rangle }{d\tau} = \langle \left [ \rholl, H_{\text{AH}}  \right ]\rangle
\label{eq_charge_eqom}
 \end{eqnarray}
 and
\begin{eqnarray}
i\hbar \frac{d \langle \mathbf{S}_{l^{\prime},l} \rangle }{d\tau} = \langle \left [ \mathbf{S}_{l^{\prime},l}, H_{\text{AH}}  \right ]\rangle,
 \end{eqnarray}
respectively.  Here the correlator:
\begin{eqnarray}
 \rholl\equiv c_{l^{\prime},\sigma}^\dagger
    c^{\phantom{\dagger}}_{l,\sigma}
 \end{eqnarray}
 is the single-particle density matrix which is off-diagonal in the site indices, $l$ and $l^{\prime}$, but measures density along the diagonal, since $\rho_{l,l}^{\sigma}=n_{l,\sigma}$.  The spin operator is:
$\mathbf{S}_{l^{\prime},l}\equiv  \psi_{l^{\prime}}^{\dagger} \boldsymbol{\sigma} \psi_{l}^{\vphantom{\dagger}}$,
where the fermion spinors are: $\psi_{l}^{\dagger}=(c^{\dagger}_{l,\uparrow},c^{\dagger}_{l,\downarrow})$ and $\boldsymbol{\sigma}$ are the Pauli matrices.  The equations of motion can be generalized to include time dependence in the Hamiltonian but we exclude that case here.

The high temperature limit studied here suppresses spin order (which emerges for $k_{B} T \lesssim t^{2}/U$).  This implies that for an equal number of atoms in each spin state we have a paramagnet:
\begin{eqnarray}
\langle  \rho_{l^{\prime}l}^{\uparrow} \rangle = \langle  \rho_{l^{\prime}l}^{\downarrow} \rangle.
 \end{eqnarray}
 To focus on the charge degrees of freedom deep in the paramagnetic regime we also assume an absence of in-plane spin order as well.  This leads to:
\begin{eqnarray}
\langle S^{x}_{l^{\prime}l}\rangle=\langle S^{y}_{l^{\prime}l}\rangle=\langle S^{z}_{l^{\prime}l}\rangle=0, 
 \end{eqnarray}
thus allowing us to focus on approximations to Eq.~(\ref{eq_charge_eqom}) only.

To derive the equations of motion we construct and solve the hierarchy of coupled differential equations with Hubbard's decoupling. 
 The commutator in Eq.~(\ref{eq_charge_eqom}) can be evaluated:
 \begin{eqnarray}
i\hbar \frac{d \langle \rholl \rangle }{d\tau} &=& (\mu_{l^{\prime}}-\mu_{l})\langle \rholl \rangle+U\langle \dll \rangle \nonumber\\
&+&\sum_{j}\left[ T_{l,j}\langle \rho_{l^{\prime}j}^{\sigma}\rangle -T_{l^{\prime},j}\langle \rho_{jl}^{\sigma} \rangle\right ]
\label{eq_eomrho}
 \end{eqnarray}
where the $U$ term contains a higher order correlator:
\begin{eqnarray}
\dll\equiv  \rholl (n_{l,-\sigma}-n_{l^{\prime},-\sigma}).
 \end{eqnarray}
The central aim of our protocol is to numerically solve Eq.~(\ref{eq_eomrho}) and use the results to evaluate Eq.~(\ref{eq_vcom}).   This will require an estimate for 
$\dll$.  

To estimate $\dll$ we derive the equations of motion for this higher order correlation function as well.  The operator evolves as:
$d\dll/d\tau=d\rholl/d\tau (n_{l,-\sigma}-n_{l^{\prime},-\sigma})+ \rholl 
(d n_{l,-\sigma}/d\tau-d n_{l^{\prime},-\sigma}/d\tau )$.  We use this relation to approximate the evolution of $\langle \dll \rangle$ by inserting 
Eq.~(\ref{eq_eomrho}) and decoupling all products of $\dll$ and $\rholl$:
 \begin{eqnarray}
&i&\hspace{-0.1cm}\hbar \frac{d \langle \dll \rangle }{d \tau} = (\mu_{l^{\prime}}-\mu_{l}) \langle \dll \rangle+U\langle \dll \rangle
\langle n_{l,-\sigma}-n_{l^{\prime},-\sigma}\rangle \nonumber \\
&+&\langle n_{l,-\sigma}-n_{l^{\prime},-\sigma}\rangle\sum_{j}\left[ T_{l,j}\langle \rho_{l^{\prime}j}^{\sigma}\rangle -T_{l^{\prime},j}\langle \rho_{jl}^{\sigma} \rangle\right ] \nonumber \\
&+&\langle \rholl \rangle\sum_{j} \left [ T_{l,j}\langle \rho_{lj}^{-\sigma}-\rho_{jl}^{-\sigma}\rangle  - T_{l^{\prime}j}\langle \rho_{l^{\prime}j}^{-\sigma}-\rho_{jl^{\prime}}^{-\sigma}\rangle\right ] 
\label{eq_eomGamma}
 \end{eqnarray}
The key decoupling used in deriving this equation is given by a Hartree-Fock-like decoupling in the equations of motion of the form:
\begin{eqnarray}
\rho_{lj}^{-\sigma} \rholl  \rightarrow \langle \rho_{lj}^{-\sigma} \rangle \rholl  \nonumber \\
  n_{l,-\sigma} \dll \rightarrow \langle n_{l,-\sigma} \rangle \dll 
  \end{eqnarray}
The next section shows that this decoupling reduces to Hubbard's decoupling \cite{hubbard:1963} that has been conventionally implemented in a Green's function approach  \cite{hubbard:1963,dirks:2014}.  It is important to note that this decoupling goes beyond conventional Hartree-Fock decouplings of the Hamiltonian \cite{penn:1966}  (which only capture the dynamics of very weakly interacting limits) to instead apply a decoupling in the equations of motion of higher order correlation functions.  It therefore offers a robust formalism that captures both weak ($t\gg U$) and strong  ($t\ll U$) interaction limits of the paramagnetic phase.

We self-consistently solve Eqs.~(\ref{eq_eomrho}) and (\ref{eq_eomGamma}) for the time evolution of the correlation functions.  We then use the correlation functions to evaluate the center-of-mass position and velocity.  The time evolution of other local correlation functions can also be found.  For example, the double occupancy, $\langle n_{l,\uparrow}n_{l,\downarrow}\rangle$, can be obtained from:
\begin{eqnarray}
 i\hbar \frac{d \langle \gammall \rangle }{d \tau} &=& (\mu_{l^{\prime}}-\mu_{l}) \langle \gammall \rangle+
 \langle \rholl \rangle \sum_{j}T_{l,j}[\langle \rho_{lj}^{-\sigma}\rangle-\langle \rho_{jl}^{-\sigma}\rangle]
\nonumber \\
 &+&\langle n_{l,-\sigma} \rangle\sum_{j}\left[ T_{l,j}\langle \rho_{l^{\prime}j}^{\sigma}\rangle -T_{l^{\prime},j}\langle \rho_{jl}^{\sigma} \rangle\right ] \nonumber \\
 &+& U \langle \gammall \rangle(1-\langle n_{l^{\prime},-\sigma}\rangle)(1-\delta_{l,l^{\prime}}) 
 \label{eomgamma}
 \end{eqnarray}
where the off-diagonal  operator:
\begin{eqnarray}
\gammall\equiv \rholl n_{l,-\sigma},
 \end{eqnarray}
 captures the conditional hopping of doublons.  

 
\section{Connection to Hubbard's Approximation}
\label{sec_hubbardapp}

In this section we argue that the formalism we have constructed can be understood in a quasiparticle picture.  
In strongly interacting systems we often rely on mappings to weakly interacting quasiparticles to gain a quantitative understanding of otherwise intractable problems.  Quasiparticles therefore offer useful tools to probe many-body localization and related phenomena.  We can then ask the following question that parallels inquiries into many-body localization of elementary particles:  Does spatial disorder localize weakly interacting quasiparticles at non-zero temperature?  Here the interactions between the original particles are strong thus allowing significant dephasing from interactions.  But quasiparticle problems are tractable and should therefore allow detailed quantitative studies.  

To connect the equations of motion to Hubbard-band quasiparticles we will show that our formalism reduces to Hubbard's approximation in the translationally invariant limit.  Our formulation is a local theory designed to incorporate spatial inhomogeneity (i.e., trapping and disorder in the quasiparticle degrees of freedom).  By assuming translational invariance we can show that the above formalism simplifies to the equations of motion found from Hubbard's approximation.  We first briefly review Hubbard's approximation.

Hubbard's approximation applies the Hartree-Fock decoupling to the equations of motion for the Green's functions.  The approximation is, unlike the ordinary Hartree-Fock approximation, exact in both the band limit (i.e., no interactions) \emph{and} the atomic limit (i.e., infinite interactions).  The approximation assumes two Hubbard bands of quasiparticles where the band parameters are renormalized by the density and the interaction.  Exact solutions of Hubbard's equations of motion are possible in the translationally invariant limit ($\mu_{j}\rightarrow -\mu_{0}$ in Eq.~(\ref{eq_muj})). The quasiparticle Green's function is:
\begin{eqnarray}
G_{\boldsymbol{k},\sigma}(E)=\frac{\hbar}{E-\left[ \varepsilon(\boldsymbol{k})-\mu_{0} +\Sigma_{\sigma}(E)\right]},
 \end{eqnarray}
where the nearest neighbor tunneling leads to the single particle band dispersion:
 \begin{eqnarray}
\varepsilon(\boldsymbol{k})\equiv -2t \sum_{\nu\in x,y,z}  \cos(k_{\nu} a).
\label{eq_nonintenergy}
 \end{eqnarray}
The self energy is \cite{hubbard:1963}:
 \begin{eqnarray}
 \Sigma_{\sigma}(E)=\frac{U\langle n_{-\sigma}\rangle(E+\mu_{0})}{E+\mu_{0}-U(1-\langle n_{-\sigma}\rangle)}.
 \end{eqnarray}
We can define a spectral density that is useful for calculations:
 \begin{eqnarray}
 \mathcal{S}_{\boldsymbol{k},\sigma}(E)=\hbar
 \delta\left[E-\varepsilon(\boldsymbol{k})+\mu_{0}-\Sigma_{\sigma}(E)\right].
 \end{eqnarray}
From the spectral density we find two (Hubbard) bands with spectral weights that depend on the density and interaction.  The energies of each band are:
  \begin{eqnarray}
  E_{b,\sigma}(\boldsymbol{k})=\frac{U+\varepsilon(\boldsymbol{k})}{2}
  +(-1)^{b}\sqrt{\frac{\left [U-\varepsilon(\boldsymbol{k})\right]^{2} }{4}+U\langle n_{-\sigma} \rangle\varepsilon(\boldsymbol{k})},
  \nonumber
  \label{eq_hubbard_energies}
  \end{eqnarray}
  where $b=1$ ($b=2$) denotes the lower (upper) Hubbard band.  In the limit of weak interaction the bands become degenerate and we recover the Hartree-Fock limit from Hubbard's approximation. 
  
\begin{figure}[t]
\includegraphics[clip,width=85mm]{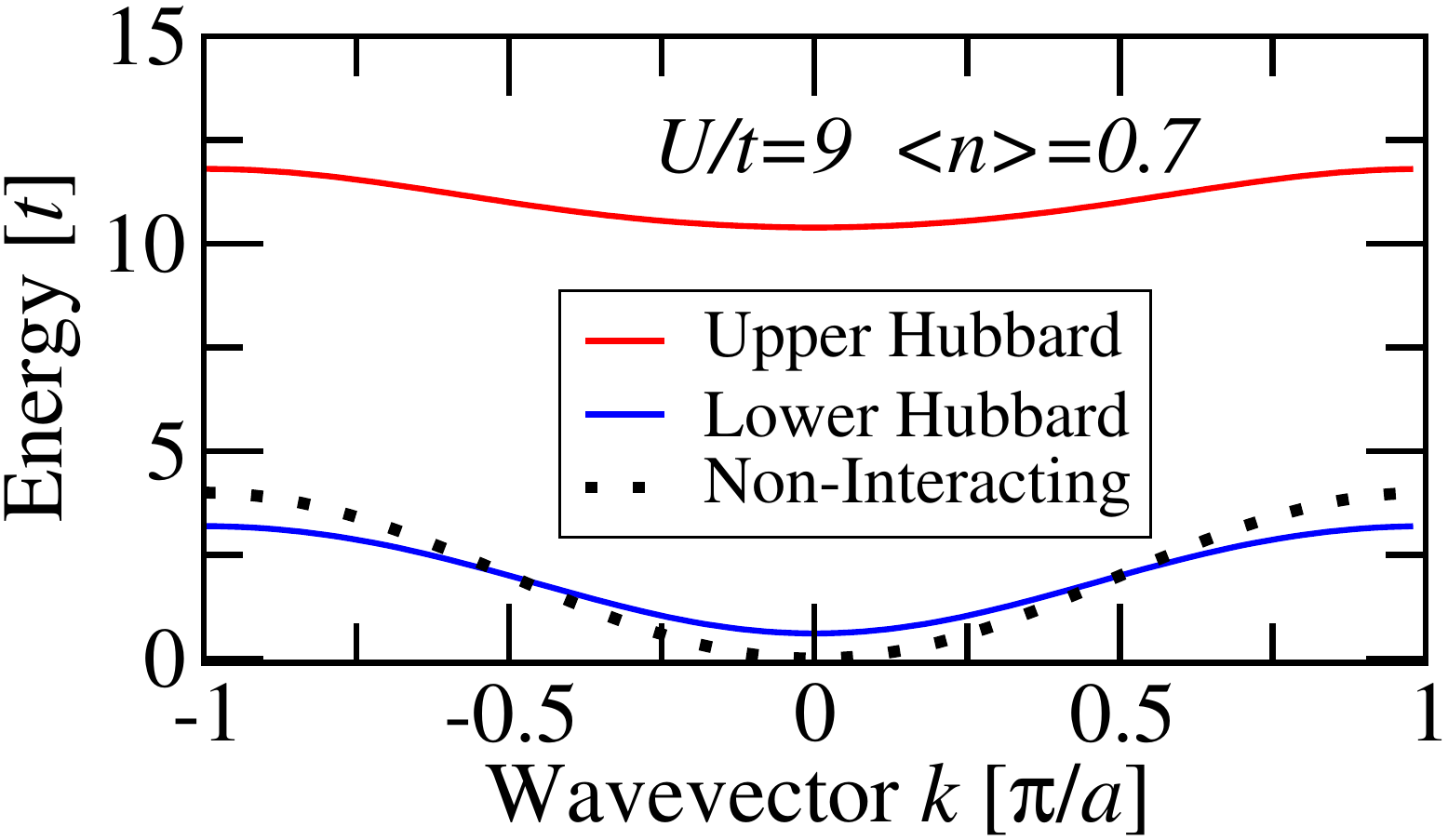}
\caption{(Color online)  Plot of the energy versus wavevector for a translationally invariant lattice.  The bottom (top) solid lines plot the energies of the lower (upper) Hubbard bands, $E_{b,\sigma}(\boldsymbol{k})$, in one dimension.  The parameters $U/t=9$ and $\langle n \rangle =0.7$ are chosen as characteristic of the center of system for the $ 7 E_{R}$ data in Table~\ref{tab1} (with $\omega=0$ and $V_{P}=0$). The dotted line plots the non-interacting case, Eq.~(\ref{eq_nonintenergy}), but in one dimension, for comparison.  Here we see that the lower Hubbard band is very similar to the non-interacting band.
}
\label{fig_hubbard_bands}
\end{figure}

  The Hubbard bands split in the strong interaction limit.  To see this we expand $E_{b,\sigma}$ in powers of $1/U$.  We find:
 \begin{eqnarray}
E_{1,\sigma}&=&  \left[1-\langle n_{-\sigma}\rangle\right ]\varepsilon(\boldsymbol{k})+\mathcal{O}(t^{2}/U) \nonumber \\
E_{2,\sigma}&=&U + \langle n_{-\sigma}\rangle\varepsilon(\boldsymbol{k})+\mathcal{O}(t^{2}/U) 
\label{eq_expandenergy}
\end{eqnarray}
This shows that, to lowest order, lower Hubbard-band quasiparticles can be thought of as non-interacting particles but with a renormalized hopping, $t \left[1-\langle n_{-\sigma}\rangle\right ]$.  (Technically, the renormalized hopping allows the quasiparticles to interact through the mean field) The upper Hubbard band is similar but with an energy offset $U$ and a renormalized hopping $t\langle n_{-\sigma}\rangle$.  An important aspect of Eq.~(\ref{eq_expandenergy}) is that the corrections are $\sim t^{2}/U$ and are therefore much smaller than the temperature in most ongoing optical lattice experiments.  Fig.~\ref{fig_hubbard_bands} plots $ E_{b,\sigma}(\boldsymbol{k})$ in comparison to $\varepsilon(\boldsymbol{k})$ along one dimension to show that the energetics of Hubbard-band quasiparticles in the lowest band are close to those of free particles. 

We now show that the formalism presented in Eqs.~(\ref{eq_eomrho})-(\ref{eomgamma}) reduces to Hubbard's approximation in the translationally invariant limit.   To show this we simplify the equations of motion for $\rholl$ and $\gammall$.  We can then solve the equations of motion by Fourier transforming into energy and wavevector variables.  We find that the resulting energies are given by $ E_{b,\sigma}(\boldsymbol{k})$.

 Eqs.~(\ref{eq_eomrho}) and (\ref{eomgamma}) define a coupled set of equations that can be solved analytically in the translationally invariant limit.  We note that these equations are coupled since $\dll =\gammall-\gamma_{ll^{\prime}}^{\sigma \dagger}$.  We impose translational invariance by setting $\mu_{l}=\mu_{l'}$.  The density then becomes uniform: $\langle n_{\sigma}\rangle=\langle n_{l,\sigma} \rangle$.  We Fourier transform all terms in Eqs.~(\ref{eq_eomrho}) and (\ref{eomgamma}).  For example, we set:
  \begin{eqnarray}
 \rhokkp=N_{s}^{-1}\sum_{l,l'}e^{-i(\boldsymbol{k}\cdot\mathbf{R}_{l}-\boldsymbol{k}'\cdot\mathbf{R}_{l'})}\rholl,
  \end{eqnarray}
 where $N_{s}$ is the number of sites.
 
 We can also transform the coupled set of first order differential equations in time to a set of coupled algebraic equations by Fourier transforming to energy space.  We then find:  
   \begin{eqnarray}
- E \rhokkp&=&U(\gamkkp-\gamkpkd)+\left [ \varepsilon(-\boldsymbol{k})-\varepsilon(\boldsymbol{k}') \right ]\rhokkp \nonumber \\
-E \gamkkp&=&U \gamkkp+\langle n_{-\sigma} \rangle \left [ \varepsilon(-\boldsymbol{k})-\varepsilon(\boldsymbol{k}') \right ]\rhokkp,
\label{eq_eqom_reduced}
 \end{eqnarray}
 where we have dropped higher order terms, i.e., terms of the form $U \langle n_{-\sigma}\rangle \gamkkp$.  We have also made use of $T_{l,l'}=N_{s}^{-1}\sum_{\boldsymbol{k}}\varepsilon(\boldsymbol{k})e^{i\boldsymbol{k}\cdot(\mathbf{R}_{l}-\mathbf{R}_{l'})}$.  
 
 Eqs.~(\ref{eq_eqom_reduced}) can be solved analytically for the eigenvalues $E$ by setting $k'=0$ and solving for $\rho_{\boldsymbol{k},0}^{\sigma}$ and $\gamma_{\boldsymbol{k},0}^{\sigma}$.  We can, without loss of generality, set $\varepsilon(0)=0$ in  Eq.~(\ref{eq_eqom_reduced}) to make contact with the Hubbard approximation. We find three distinct modes.  A trivial high energy mode with $E=U$ corresponds to a non-dispersive doublon mode obtained from solutions with $\rho_{\boldsymbol{k},0}^{\sigma}=0$.  But the remaining two modes we find have precisely the same energies as those found in Hubbard's approximation: $E_{b,\sigma}(\boldsymbol{k})$.  We have therefore shown that the 
formalism presented in Eqs.~(\ref{eq_eomrho})-(\ref{eomgamma}) reduces to Hubbard's approximation in the translationally invariant limit. 
 
The reduction of the transport problem posed by Eq.~(\ref{eq_hubbardmodel}) in a high temperature paramagnetic limit into that of transport of Hubbard-band quasiparticles has two important implications.  The first is practical: We will be able to compute correlation functions for the initial state using the spectral density.  This is discussed in Sec.~\ref{sec_initial_state}.  

The second implication is phenomenological.  Since the strongly interacting limit can be thought of as nearly free quasiparticles, the addition of disorder should show features qualitatively similar to a weakly interacting system.  We have verified that the quasiparticle picture remains valid even for large disorder strengths, $\Delta_{E}/t\sim40$, by checking that the Hubbard-band spectral weight is non-zero.  We can therefore construct an effective model of quasiparticles in a disordered lattice (but in the absence of a trap):
 \begin{eqnarray}
 H_{\text{eff}}= \sum_{j,j',\sigma,b} \tilde{T}_{j,j'}^{b,\sigma} \tilde{c}_{j,b,\sigma}^\dagger
    \tilde{c}^{\phantom{\dagger}}_{j',b,\sigma}
     +\sum_{j,b,\sigma} \tilde{\mu} _{j,b,\sigma}  \tilde{n}_{j,b,\sigma}
     \label{eq_heff}
 \end{eqnarray}
where the tilde indicates quasiparticle operators.  $\tilde{\mu}$ is the chemical potential renormalized by the self energy and $ \tilde{T}_{j,j'}$ indicates quasiparticle hopping with:
 \begin{eqnarray}
 \tilde{T}_{j,j'}^{b,\sigma}=N_{s}^{-1}\sum_{\boldsymbol{k}}E_{b,\sigma}(\boldsymbol{k})e^{i\boldsymbol{k}\cdot(\mathbf{R}_{l}-\mathbf{R}_{l'})}.  
\end{eqnarray}
Here we have assumed that the quasiparticle energies, $E_{b,\sigma}(\boldsymbol{k})$, depend on the Fourier transform of the randomly distributed density. 

\begin{figure}[t]
\includegraphics[clip,width=80mm]{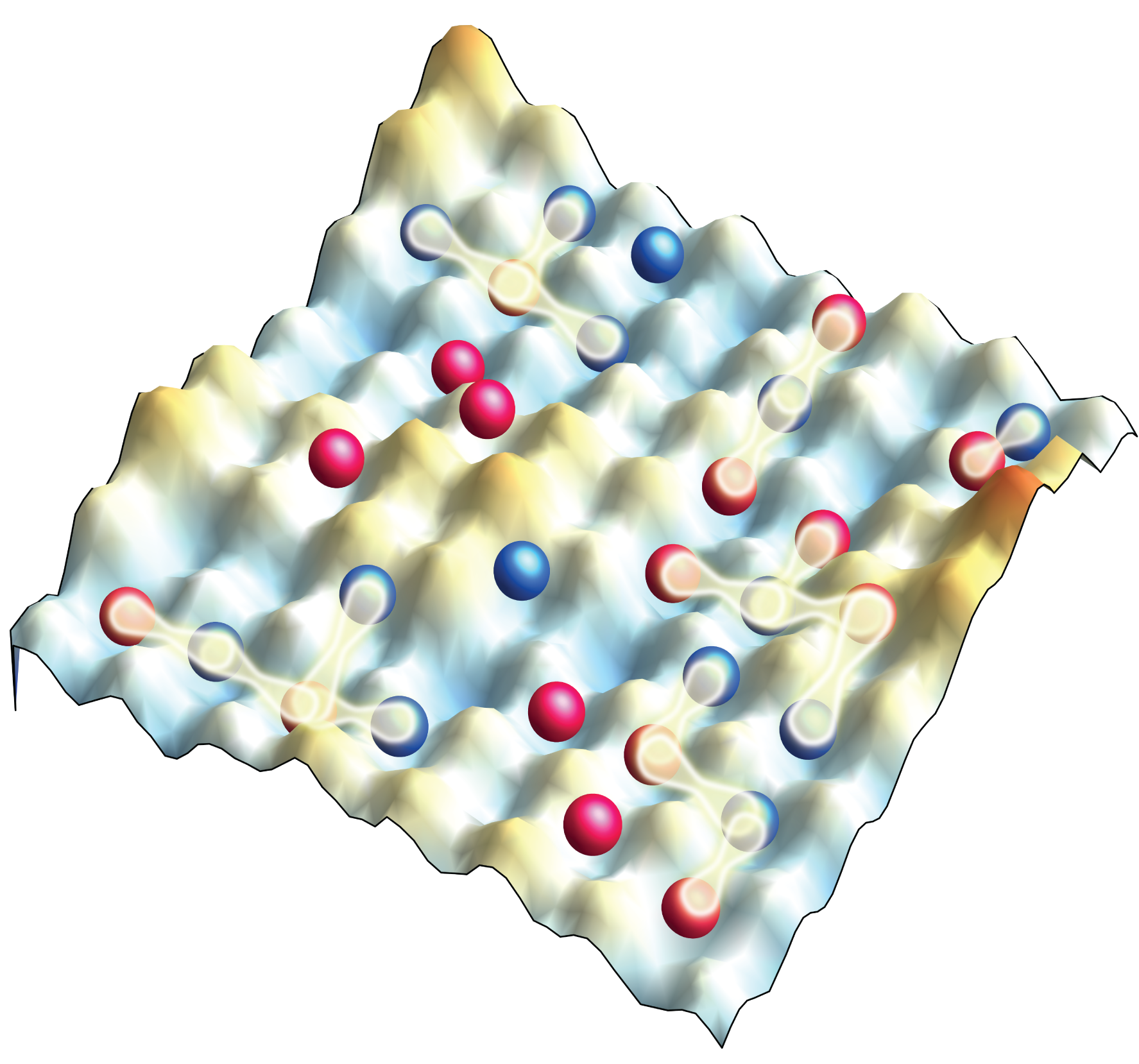}
\caption{(Color online)  Schematic of one two-dimensional plane of the site-disordered cubic lattice.  The blue (red) spheres depict spin up (down) quasiparticles in the lowest Hubbard band.  The highlighted nearest neighbor bonds between particles of opposite spin symbolize the renormalized quasiparticle hopping in Eq.~(\ref{eq_heff}).  The hopping is suppressed on average. For example, a lower-band quasiparticle with spin $\sigma$ has a renormalize nearest neighbor hopping $t\langle 1-n_{-\sigma} \rangle$, for large $U$. }
\label{fig_schematic_quasiparticle}
\end{figure}

We can get an intuition for the renormalized hopping if we assume that the density, on average, remains uniform in the presence of disorder.  Eqs.~(\ref{eq_expandenergy}) show that in the strongly interacting limit this renormalized hopping reduces to:  $ \tilde{T}_{j,j'}^{1,\sigma}\approx T_{j,j'} \langle 1-n_{-\sigma} \rangle+\mathcal{O}(t^{2}/U)$ and $ \tilde{T}_{j,j'}^{2,\sigma}\approx  T_{j,j'} \langle n_{-\sigma} \rangle +\mathcal{O}(t^{2}/U)$, for the lower and upper Hubbard bands, respectively.  The renormalized hopping is shown schematically in Fig.~\ref{fig_schematic_quasiparticle}.   

$H_{\text{eff}}$ is an effective theory of quasiparticles that must, in principle, be solved self-consistently.  But it should nonetheless reveal a mobility edge of Hubbard-band quasiparticles because it is essentially a non-interacting Anderson model of Hubbard-band quasiparticles.  For example, it is well known that the Anderson model in the cubic lattice with uniform disorder exhibits a mobility edge at $\Delta_{U}/B=X_{c}$, where $X_{c}\approx1.6$ \cite{bulka:1987}.  For the lower Hubbard band in the absence of a trap and in a paramagnetic state the quasiparticle bandwidth becomes $B=12t (1- \langle n \rangle/2)$.  $H_{\text{eff}}$ therefore qualitatively predicts a mobility edge for Hubbard-band quasiparticles.  We will return to this point in discussing the suppression of transport in Sec.~\ref{sec_discussion}.  
 
$H_{\text{eff}}$ also shows that  ignoring disorder in $t$ and $U$ is justified in the large $U$ limit.  It Ref.~\cite{zhou:2010} it was shown that speckle disorder leads to a narrow Lorentzian distribution of $t$ and $U$.  Even though the distribution is narrow, these parameters could in principle have significant contributions to transport due to the tails of the distribution.  But we note that the large $U$ limit is dominated by transport of quasiparticles (not the original particles).  Eqs.~(\ref{eq_expandenergy}) and ~(\ref{eq_heff}) explicitly show that the effective quasiparticle hopping, $\tilde{T}$, and chemical potential, $\tilde{\mu}$, are implicitly disordered even if disorder in $t$ and $U$ are excluded. This shows that excluding disorder in $t$ and $U$ still leaves an effective model with all terms disordered.  Including disorder in $t$ and $U$ should therefore not qualitatively impact the transport properties of quasiparticles in the large $U$ limit.

\section{Initial State}
\label{sec_initial_state}

To time evolve correlation functions we must accurately establish the initial state.  The system evolves in the absence of a heat or particle number bath.  The short-time dependence therefore crucially depends on the initial state.  We note that the Hubbard approximation is very accurate in the limit defined by Eqs.~(\ref{eq_approx}).  To see this we note that the static properties of optical lattice experiments with two-component fermions are also accurately captured by a high temperature series expansion of Eq.~(\ref{eq_hubbardmodel}) \cite{oitmaa:2006,scarola:2009a,deleo:2011}.  We have checked that the high temperature series expansion and the Hubbard approximation agree in the limits discussed here, Eqs.~(\ref{eq_approx}).
  
In this high $T$ regime the initial state is also accurately captured by the local density approximation \cite{scarola:2009a,dirks:2014}.  We take each site as a uniform system and compute correlation functions.  In the local density approximation we assume that each site of the trapped system can be approximated with parameters for a uniform system by setting $\mu_{l}$ to be the chemical potential for the $l^{th}$ uniform system, and we average over all uniform systems.  Correlation functions from each site are then combined. In the case of multi-site correlation functions a complication arises: the chemical potential varies from site to site.  Here we find that nearest neighbor correlation functions are sufficient to describe the initial state, since long range correlation functions decay quickly at these temperatures.  As a result, we are able to approximate two-site correlation functions by setting the chemical potential to be the average between the neighbors.

The initial state can be approximated using correlation functions computed directly from the spectral function within the local density approximation.  The spectral theorem implies that we can compute the initial ($\tau=0$) correlation function for $\rholl$ using:
\begin{eqnarray}
\langle \rholl \rangle(\tau=0)
&=&\sum_{\boldsymbol{k}}\frac{e^{-i(\boldsymbol{R}_{l'}-\boldsymbol{R}_{l})\cdot\boldsymbol{k}}}{2\hbar N_{s}} \int_{-\infty}^{\infty}dEf(E) \nonumber \\
&\cdot& \mathcal{S}_{\boldsymbol{k},\sigma}(E-\bar{\mu}),
\label{eq_initrho}
 \end{eqnarray}
 where $\bar{\mu}\equiv (\mu_{l}-\mu_{l'})/2$ and $f(E)$ is the Fermi-Dirac distribution function.  Here we assume $\langle \rholl \rangle$ is equal to its Hermitian conjugate.  A similar relation can be used to obtain $\dll$ as well:
\begin{eqnarray}
\langle \rholl n_{l,-\sigma} \rangle(\tau=0) &=& \sum_{\boldsymbol{k}}\frac{e^{-i(\boldsymbol{R}_{l'}-\boldsymbol{R}_{l})\cdot\boldsymbol{k}}}{2\hbar N_{s}} \int_{-\infty}^{\infty}dEf(E) \nonumber\\
&\cdot&
\left [ \frac{E-\varepsilon(\boldsymbol{k})}{U} \right ]\mathcal{S}_{\boldsymbol{k},\sigma}(E-\bar{\mu}).
\label{eq_initgamma}
 \end{eqnarray}

Using the these relations we are able to set the initial state correlation functions with a protocol discussed in Sec.~\ref{sec_com_dynamics}.  The protocol allows use of the Hubbard approximation to compute initial state correlation functions at fixed entropy for a given disorder configuration.  The following discusses the temperature dependence in the initial state in the presence of disorder.

\section{Adiabatic Heating due to Disorder In the Initial State}
\label{sec_heating}

The temperature in ultracold atom experiments is determined by the entropy.  The relationship between temperature and entropy relies, in general, on the intricate interplay between kinetics and interactions.  The addition of disorder adds another complication that alters the entropy-temperature relation.  Below we show that the addition of disorder leads to adiabatic heating in the initial state.  Specifically we find that, at fixed entropy, increasing disorder increases the temperature.  This observation has important consequences for the interpretation of the data in Ref.~\cite{kondov:2015} and other optical lattice experiments because increasing disorder strengths also increases temperature.  In subsequent sections we take adiabatic heating from disorder into account when preparing the initial state in a trap.

\begin{figure}[t]
\includegraphics[clip,width=85mm]{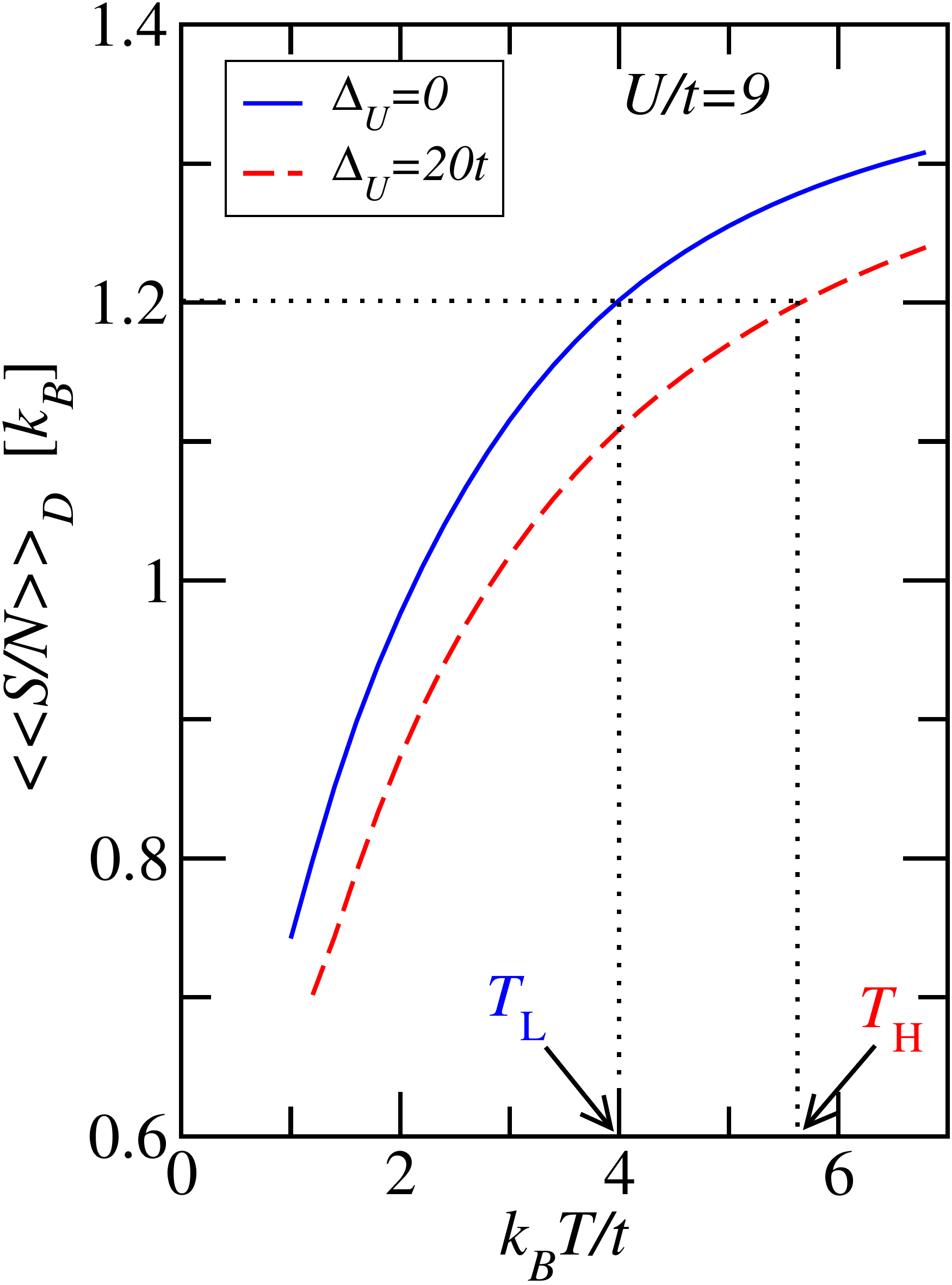}
\caption{(Color online)  The disorder-averaged entropy per particle computed as a function of temperature for Eq.~(\ref{eq_hubbardmodel}) in the absence of a trapping potential or a pulse ($\omega=0$ and $V_{P}=0$).  The horizontal dotted line indicates a fixed entropy per particle, $S/N=1.2$$k_{\text{B}}$.  The solid (dashed) lines were computed using $\Delta_{U}=0$ ($\Delta_{U}=20t$) and $\mu_{0}/t=3.8$.  The vertical lines labeled with $T_{L}$ and $T_{H}$ point to low and high temperatures, respectively.  The entropy-temperature curve with a high disorder leads to a higher temperature.
}
\label{fig_entropy_2D}
\end{figure}

We use the high temperature series expansion to show that the paramagnet experiences adiabatic heating due to disorder.  The solid line in Fig.~\ref{fig_entropy_2D} shows an example of the entropy per particle versus temperature for an initial state without a trap.   We set $\mu_{0}/t=3.8$ because it characterizes the non-disordered limit of experiments reported in Ref~\cite{kondov:2015}.  We find $\langle n \rangle <1$.   Here we see that a fixed entropy (horizontal dotted line) sets a low temperature, $T_{L}$, in the absence of disorder.  Because optical lattice experiments take place in the absence of a heat bath, entropy is preserved when a disordered optical lattice is applied to a trapped gas.  We then include a disorder strength $\Delta_{U}=20t$ in a calculation of the entropy per particle.  We use the local density approximation and integrate over disorder configurations (See Eqs.~(\ref{eq_denent})).  The dashed line shows the disorder averaged results.  The entropy is significantly lower.  The system therefore acquires a higher temperature, $T_{H}$, at the same entropy. 

Adiabatic heating due to disorder arises  because increasing the disorder strength in a single band reduces the number of available states.  As a result the entropy (which is the logarithm of the number of available states) decreases with increasing disorder.  The net effect is then an increase in temperature if the entropy is required to be fixed while increasing disorder. 

\begin{figure}[t]
\includegraphics[clip,width=85mm]{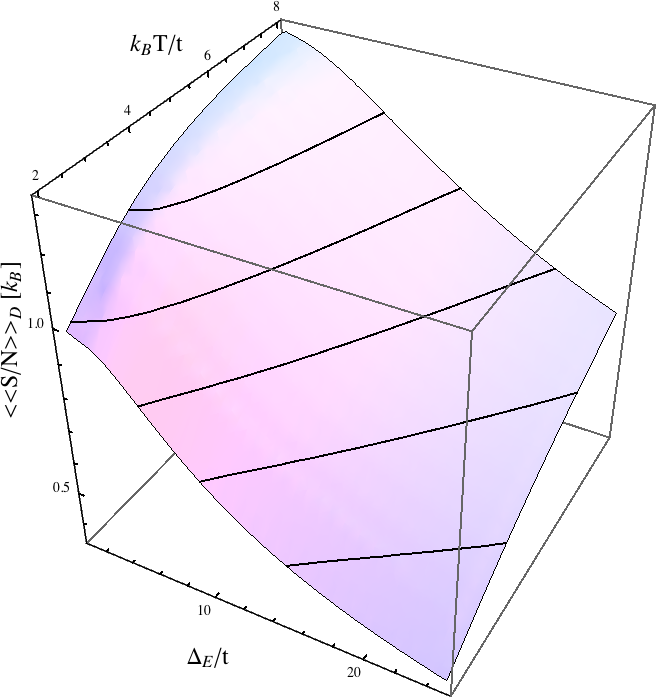}
\caption{(Color online)  Disorder averaged entropy per particle plotted as function of both temperature and disorder strength for Eq.~(\ref{eq_hubbardmodel}) in the absence of a trapping potential or an applied force (i.e., $\omega=0$ and $V_{P}=0$).  The $8^{th}$ order high temperature series expansion was used within the local density approximation.  $U/t=9$ and $\mu_{0}/t=3.8$ were chosen as characteristic of the center of system for the $ 7 E_{R}$ data in Table~\ref{tab1}.  The black contour lines indicate adiabats that reveal significant adiabatic heating due to increasing exponential disorder. }
\label{fig_entropy_3D}
\end{figure}

Adiabatic heating becomes more pronounced with exponentially distributed disorder.  Fig.~\ref{fig_entropy_3D} plots the entropy per particle as a function of both exponential disorder strength and temperature.  The black lines depict adiabats.  The corresponding temperature can therefore increase by as much as a factor of 2 at fixed entropy over the range of disorder strengths considered here.  The impact of adiabatic heating due to disorder on center-of-mass transport in trapped systems is discussed in more detail in the following sections.

\section{center-of-mass Dynamics: Comparison with Experiment}
\label{sec_com_dynamics}

This section culminates in a direct comparison between results from the equations of motion and experiments.  We find that small system size simulations can be scaled to directly compare with experiments with no fitting parameters.  The close comparison between experiment and theory shows that we can interpret the experiments of Ref.~\cite{kondov:2015} as transport of Hubbard-band quasiparticles.  The simulations and experiments are consistent with finite size precursors of Anderson localization of Hubbard-band quasiparticles.

We now use Eqs.~(\ref{eq_initrho}) and (\ref{eq_initgamma}) to compare with experiments in Ref.~\cite{kondov:2015} using experimental input parameters from Table~\ref{tab1}.  To use our formalism to compute the center-of-mass dynamics we prepare an initial state at fixed entropy in a disordered landscape.  The system is numerically time evolved.  The center-of-mass velocity is computed at the pulse time and then disorder averaged.  These simulations are performed on system sizes up to $L=11$, with $L=L_{x}=L_{y}=L_{z}$.  Finite size extrapolation is performed by decreasing the trap frequency and repeating the simulation for large system sizes while keeping $\mu_{0}$ fixed to values found for experimentally relevant system parameters.  

To keep the pulse time short on the time scales of the trapping frequency (as is done experimentally \cite{kondov:2015}) we have to rescale the pulse time used in our simulations.  The pulse time at system size $L$, $\tau_{L}$, is adjusted for each trap frequency at system size $L$, $\omega_{L}$, to maintain $\tau_{L}=\tau_{P}\sqrt{\omega/\omega_{L}}$.  This allows a scaling to the trapping frequency and the pulse time found in Table~\ref{tab1}, $\omega$ and $\tau_{P}$, respectively.  The impulse formula (See Sec.~\ref{sec_order}) shows that this establishes an $\omega_{L}^{-1/2}$ scaling of $V_{\text{C.O.M.}}$.  This scaling is expected since the center-of-mass velocity from the impulse formula scales as $V_{0}\sim\tau_{L}\sim{\omega_{L}^{-1/2}}$ (see Sec.~\ref{sec_order}).  We have checked below that our finite size extrapolations do scale as $\omega_{L}^{-1/2}$, as expected.   

\begin{figure}[t]
\includegraphics[clip,width=85mm]{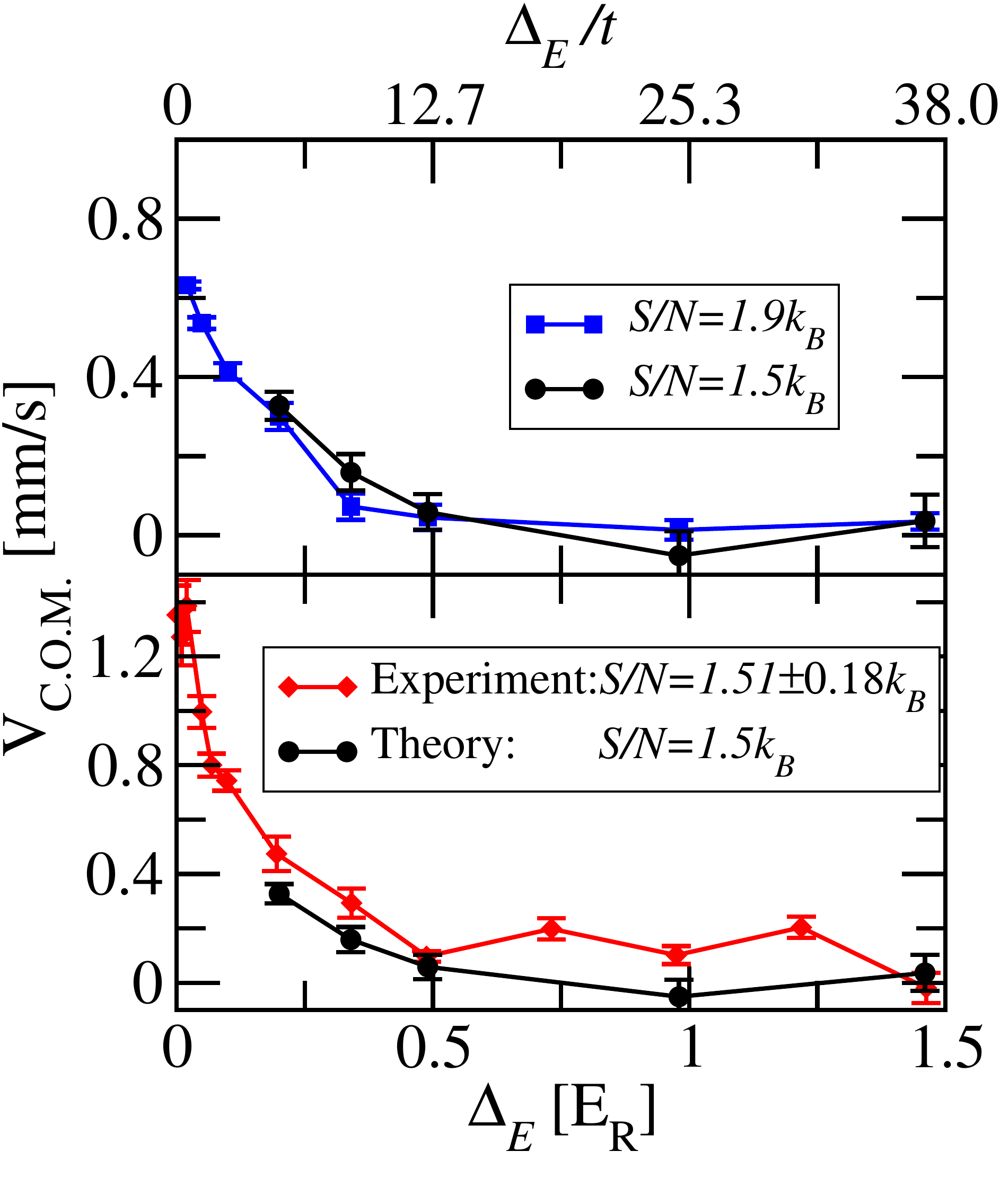}
\caption{(Color online) Top: Disorder averaged center-of-mass velocity as a function of the disorder strength for two different entropies.  Here the initial state correlation functions are estimated in the local density approximation (using Eqs.~(\ref{eq_initrho}) and ~(\ref{eq_initgamma}) in combination with a high temperature series expansion) and time evolved in the trap (using Eqs.~(\ref{eq_eomrho}) and (\ref{eq_eomGamma})).  Model parameters are taken from the $6 E_{R}$ data in Table~\ref{tab1}.  The $S/N=1.5 k_{B}$ results are  plotted only for large disorder strengths because here adiabatic heating allows access to temperatures high enough to be consistent with the approximations made in preparing the initial state.  Bottom: The circles plot the same as the top panel and  the diamonds plot experimental data from Ref.~\cite{kondov:2015} for comparison.  The lines are a guide to the eye.  The error bars on the numerical simulations are the standard error found from disorder averaging, while the experimental error bars are the standard error in the mean for 7-9 measurements averaged for each point.
}
\label{fig_comparison_6ER}
\end{figure}

\begin{figure}[t]
\includegraphics[clip,width=85mm]{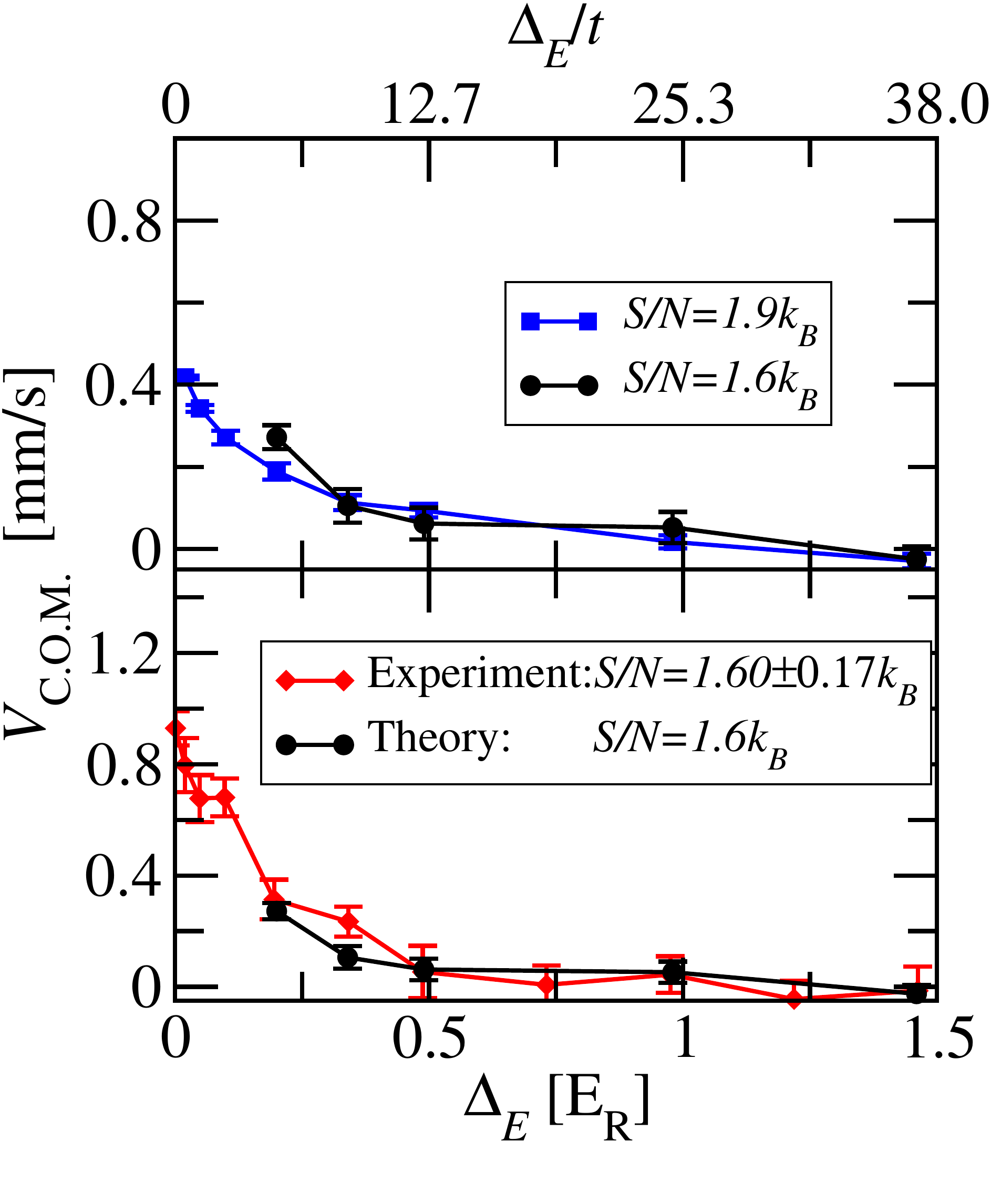}
\caption{(Color online)  The same as Fig.~\ref{fig_comparison_6ER} but for the $7 E_{R}$ data in Table~\ref{tab1}.  Here the comparison between theory and experiment is better because $U/t$ is larger.}
\label{fig_comparison_7ER}
\end{figure}

We use the following protocol to prepare initial states:  1)  We choose an entropy per particle determined experimentally, high trap frequency (chosen to trap the system within the finite size limitations of our simulations), and a small number of particles. 2) We choose a random distribution of chemical potentials according to Eq.~(\ref{eq_PE}).  3) We then self-consistently adjust $\mu_{0}$ and $T$  so that the particle number and entropy match the values set in step 1.  This is done using a high temperature series expansion in  the local density approximation.  The series expansion is controlled a these temperatures because we can check higher orders \cite{scarola:2009a,deleo:2011}.   We find that $8^{th}$ order in the expansion is sufficient for parameters considered here.  The Hubbard approximation gives identical results for thermodynamic functions.  4) We then use Eqs.~(\ref{eq_initrho}) and (\ref{eq_initgamma}) to compute the initial state correlation functions.  5)  We then return to step 1 to repeat the process with a smaller trap frequency.

We find that adiabatic heating in the initial state increases the temperature by no more than a factor of 2.  For all system sizes studied we find that the temperature remains nearly constant as function of system size.  At the largest disorder strengths, $\Delta_{E}\sim 1.5 E_{R}$, we still find $k_{B}T<4t$.  We conclude that adiabatic heating increases the temperature but the temperature is still well below the bandwidth, $12t$.

Given the initial state, we numerically time evolve correlation functions according to Eqs.~(\ref{eq_eomrho}) and Eqs.~(\ref{eq_eomGamma}), extrapolate to the thermodynamic limit, and disorder average.  Figs.~\ref{fig_comparison_6ER} and \ref{fig_comparison_7ER} plot $V_{\text{C.O.M.}}$ versus disorder strength for the $6$ $E_{R}$ and the $7$ $E_{R}$ parameters, respectively.  The data result from time evolving the initial correlators, Eqs.~(\ref{eq_initrho}) and (\ref{eq_initgamma}).  The top panels show results for two different entropies.  The larger entropy leads to temperatures with $T\gtrsim t$.  The approximations made here (paramagnetic order, no spin correlations, and the local density approximation) are therefore valid at all disorder strengths for the higher entropy.  The top panels also compare low entropy data that is consistent with the entropies used in experiments (see Table~\ref{tab1}).  Here adiabatic heating increases the temperature to $T\gtrsim t$ only for $\Delta_{E} \gtrsim 0.2 E_{R}$.  Below these disorder strengths the approximations made here break down because the temperatures are low enough to introduce poles in thermodynamic functions using either the high temperature series expansion (even out to $10^{th}$ order) or the Hubbard approximation.

The top panels of Figs.~\ref{fig_comparison_6ER} and \ref{fig_comparison_7ER} clearly show a suppression of the center-of-mass velocity with disorder.  The mapping to quasiparticles in the lowest Hubbard band allows delineation of the sources of the suppression: 1) As exponentially distributed disorder is increased, the bias in the distribution leads to more sites with higher densities.   The increase in average density slows the propagation of the Hubbard-band quasiparticles because the renormalized tunneling is given by $t\langle 1- n/2\rangle$.  This effect was implicit in the suppression shown in Sec.~\ref{sec_order} (See Fig.~\ref{fig_impulse}). We find that this is a weak effect because the system is dilute, i.e., $\langle n_{j} \rangle/2 \ll 1$ for many sites (the edges make up about 1/3 of the system) 2) Adiabatic heating due to disorder also suppresses $V_{\text{C.O.M.}}$.  The increase in the resulting temperature lowers the nearest neighbor correlations, e.g., $\langle \rho^{\sigma}_{l,l+1} \rangle$, inherent in the initial state.  The initial state is therefore slower to respond because $V_{\text{C.O.M.}}$ scales linearly with terms like  $t\langle \rho^{\sigma}_{l,l+1} \rangle$.  This effect was shown to dominate only at lower disorder in Sec.~\ref{sec_temp} (See Fig.~\ref{fig_v_vs_T}).  Furthermore, we find that the temperature is at most $B/3$ at the largest disorder strength, $\Delta_{E}\sim 1.5 E_{R}$. 3)  These effects are modest and are not sufficient to completely localize the center of mass.  The final effect derives from disorder induced scattering.  The presence of disorder lowers the localization length so that propagation is impossible for $\Delta_{E} >0.5 E_{R}$.  This final effect is consistent with a finite size precursor of Anderson localization of Hubbard-band quasiparticles because the critical disorder strength, $\Delta_{E} \approx 0.5 E_{R}$, is near the approximate location expected for the Anderson metal-insulator transition, near $B\approx 0.47 E_{R}$.

The bottom panels in Figs.~\ref{fig_comparison_6ER} and \ref{fig_comparison_7ER} show a comparison between the results obtained from our formalism and the experimental data of Ref.~\cite{kondov:2015}.  The comparison is made where possible (in the high temperature regime).  The agreement in Fig.~\ref{fig_comparison_7ER} is better because $U$ is larger.  The Hubbard approximation is technically a strong coupling approximation that is exact in the $U/t\rightarrow\infty$ limit.  The comparison suggests that the data from Ref.~\cite{kondov:2015} can be thought of as revealing a mobility edge of Hubbard-band quasiparticles.  

\section{Discussion}
\label{sec_discussion}

We have found that two-component fermions in an optical lattice fail to respond to a force and undergo mass transport for sufficiently strong disorder, implying a phenomenon reminiscent of Anderson localization in bulk systems.  At strong disorder strengths the atoms fail to move under weak perturbations.  Here the suppression of quantum diffusion indicates that the assumption of a thermal initial state is incorrect, i.e., that the system is inherently non-ergodic at large disorder strengths.  Our comparison between theory and experiment are therefore consistent with Anderson localization of Hubbard-band quasiparticles at large disorder strengths but a mobile state of Hubbard-band quasiparticles at low disorder strengths.  We interpret these results as evidence for a mobility edge of Hubbard-band quasiparticles.

We can compare  the center-of-mass velocity studied here with conductivity studied in solids.  Both measures can be used as diagnostics of localization.  The DC Conductivity in solids is typically defined in infinite system sizes.  The DC conductivity therefore gives a long time/large length scale probe of the single particle density matrix.  The center-of-mass velocity is proportional to mobility and therefore also offers an equivalent probe of the single particle density matrix provided the system is infinitely large and it is allowed to evolve indefinitely.  But the center-of-mass velocity studied here was considered on time scales inversely proportional to the trap frequency and in finite system sizes.  We therefore conclude that the results presented in Figs.~\ref{fig_comparison_6ER} and \ref{fig_comparison_7ER} only offer a finite size estimate for the conductivity.

Our work opens interesting directions for future studies of localization physics with Hubbard-band quasiparticles.  The work presented here is consistent with quantum Monte Carlo results \cite{denteneer:1999} and dynamical mean field theory studies of the Anderson-Hubbard model \cite{georges:1996,byczuk:2005,semmler:2010}.  But these methods could be used to tackle lower temperature limits and include spin fluctuations in a comparison with low temperature experiments.  

Furthermore, future work will be needed to rigorously establish a connection between the localized state found here and many-body localization.  The suppression of transport at non-zero temperatures found here is a necessary condition for many-body localization.  But future work should look at sufficient conditions for many-body localization using, e.g., entanglement measures in the Anderson-Hubbard model, to make a direct comparison with experiments. 

In preparing this manuscript we became aware of work in Ref.~\cite{schreiber:2015} that compared the entanglement entropy with population imbalance in incommensurate optical lattices.

 V.W.S. acknowledges support from AFOSR under grant FA9550-11-1-0313.  B.D. acknowledges support from the NSF under grant PHY12-05548 and from the ARO under grant W9112-1-0462.

\appendix

\section{Order of Magnitude Estimate}
\label{sec_order}

This section uses a semiclassical impulse formula for Hubbard-band quasiparticles to estimate the center-of-mass velocity dependence on disorder strength for very weak disorder.  This estimate shows that renormalization of the quasiparticle hopping due to disorder can suppress the center-of-mass velocity.  It also yields the correct order of magnitude for the center-of-mass velocity at low disorder.  A simple order of magnitude estimate for the center-of-mass velocity will be useful in establishing a scaling relation to extrapolate our finite sized simulations to experimental system sizes.  

To estimate the center-of-mass velocity we use a semiclassical estimate of velocities in combination with the local density and effective mass approximations.  The quasiparticle effective mass in the lowest Hubbard band is obtained from the single-particle effective mass using the replacement $t\rightarrow t\langle 1- n_{j}/2\rangle$:
\begin{eqnarray}
m^{*}_{j}=\frac{\hbar^{2}}{2 t( 1- \langle \langle n_{j}\rangle \rangle_{D}/2) a^{2}},
\label{Eq_qpeffmass}
 \end{eqnarray}
 where the limit $\langle n_{j} \rangle\rightarrow 0$ returns the single-particle effective mass.  Note that disorder averaging is implicit in this definition.   

At short times, the semiclassical estimate of the center-of-mass velocity reduces to the well known impulse formula. We apply the impulse formula to the dynamics of Hubbard-band quasiparticles in the lowest band.  (Note that the impulse formula also follows from the generalized Kohn's theorem in an effective mass approximation) Averaging the velocity of each site, $\dt{\langle R_{j}^{x} \rangle}$, leads to a total center-of-mass velocity for one disorder configuration $N^{-1}\sum_{j}^{N_{s}}\langle n_{j} \rangle \dt{\langle R_{j}^{x} \rangle}$.  Applying the impulse formula to Hubbard-band quasiparticles and averaging over disorder realizations gives an approximation to the center-of-mass velocity: 

\begin{eqnarray}
V^{\text{I}}=V_{0}[1-\sum_{j}^{N_{s}}\langle \langle n_{j} \rangle^{2}\rangle_{D} /(2N)],
\label{eq_impulse}
 \end{eqnarray}
 where $V_{0}\equiv2aV_{P}\tau_{P}t/\hbar^{2}$ depends linearly on $\tau_{P}$ and $\langle \langle n_{j} \rangle^{2}\rangle_{D}$ indicates the disorder average of $\langle n_{j} \rangle^{2}$.  

$V^{\text{I}}$ gives the correct order of magnitude for the center-of-mass velocity.  To show this we use the high temperature series expansion to estimate the density in the initial state in the trap.  We choose the parameters for the $7 E_{R}$ lattice depth presented in Table~\ref{tab1} but we fix the entropy to be $S=1.9 k_{B}$.  

\begin{figure}[t]
\includegraphics[clip,width=80mm]{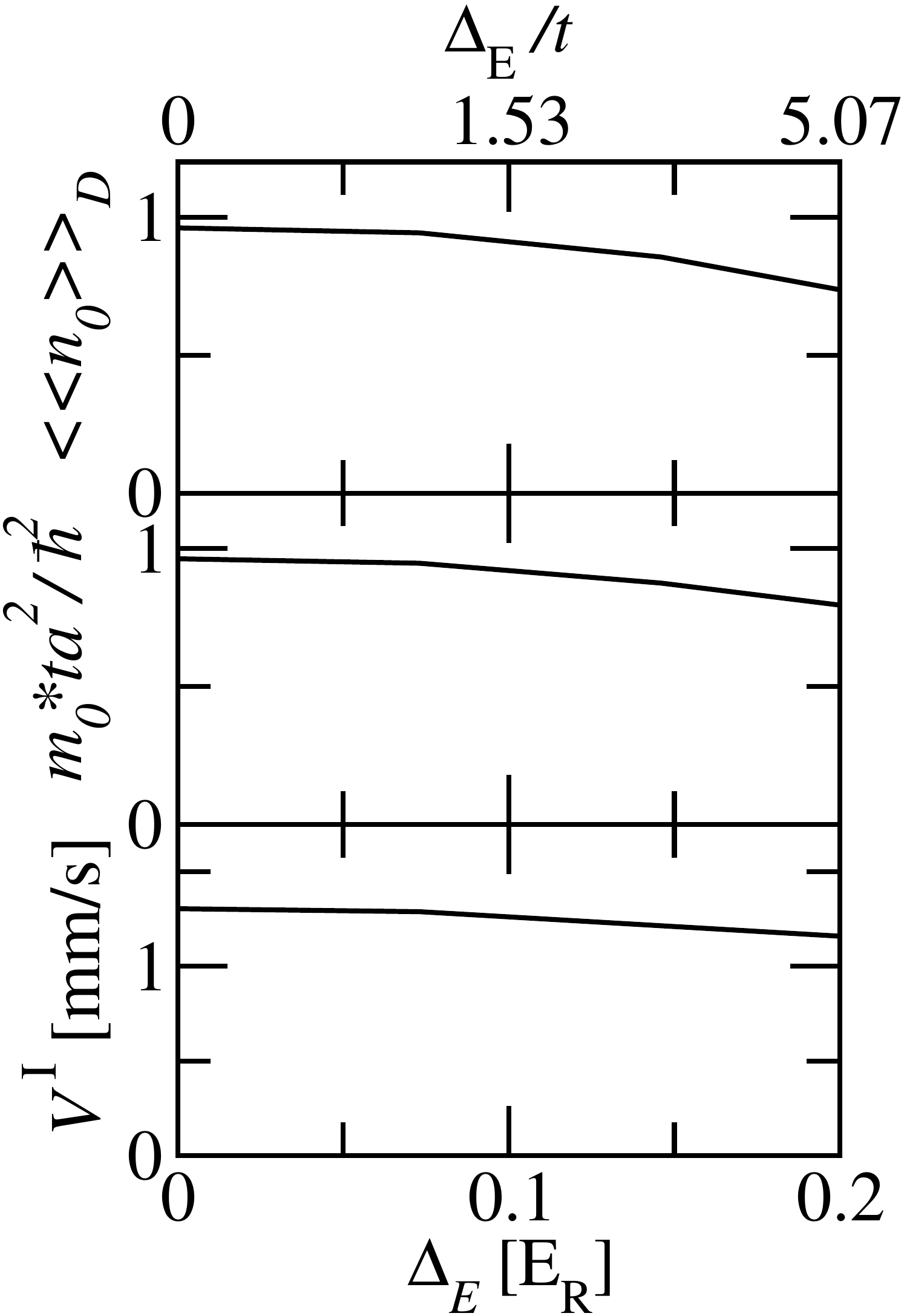}
\caption{(Color online)  The top and middle panels plot the density and the quasiparticle effective mass, Eq.~(\ref{Eq_qpeffmass}), respectively, as a function of disorder strength for a chemical potential at the central site.  The bottom panel plots the center-of-mass velocity versus disorder strength from a disorder-averaged impulse formula, Eq.~(\ref{eq_impulse}), that estimates the velocity of Hubbard-band quasiparticles in the trap size consistent with experiment.   The local density approximation was used to sum over all sites.  All quantities are computed using the high temperature series expansion at $8^{th}$ order with the parameters chosen from the $7 E_{R}$ data in Table~\ref{tab1} but with $S/N=1.9k_{B}$.  Eqs.~(\ref{eq_denent}) were used as rough estimates for disorder averaging. }
\label{fig_impulse}
\end{figure}

We use a simplified version of the protocol constructed in the main text to get a rough estimate of $V^{\text{I}}$.  
Once the entropy and particle number are fixed, the approach used in the main text then finds the $\mu_{0}$ and $T$ at \emph{each} disorder configuration using the high temperature series expansion.  These parameters are then, for each disorder configuration, used to compute $\langle n_{j} \rangle$ within the trap.  Disorder averaging proceeds by summing the center-of-mass velocity over all disorder configurations.  But in this section we solve for the chemical potential and temperature differently so we can access experimentally relevant system sizes without finite size extrapolation.  We use the high temperature series expansion to approximate the entropy and density with integration (rather than explicit summation) over the  disorder distribution:
\begin{eqnarray}
 \langle\langle S \rangle \rangle_{D} &\approx& \int_{0}^{\infty} d\epsilon P_{E}(\epsilon) S(\epsilon) \hspace{0.5cm} \nonumber \\
 \langle\langle n \rangle \rangle_{D} &\approx& \int_{0}^{\infty} d\epsilon P_{E}(\epsilon) \langle n(\epsilon) \rangle.
 \label{eq_denent}
 \end{eqnarray}
 These approximations can be used to self-consistently solve for $T$ and $\mu_{0}$ given $S$ and $N$ for large systems sizes.  This simplified protocol uses entropies and densities that are not self-consistently solved for each disorder configuration but are instead taken in a mean-field limit separately.  As a result, self-consistent solutions of these coupled formulas only offer a rough estimate for $T$ and $\mu_{0}$ because they are assumed to decouple for each disorder configuration.  We can therefore only apply these approximations for low disorder strengths.

The top panel of  Fig.~\ref{fig_impulse} plots the disorder-averaged density of the central site in the trap as a function of disorder strength.  Here we see that the density decreases due to adiabatic heating and a redistribution of the particles due to biased exponential disorder.  The quasiparticle effective mass (middle panel) therefore also decreases.  

The bottom panel of Fig.~\ref{fig_impulse} plots the disorder-averaged center-of-mass velocity from Eq.~(\ref{eq_impulse}).  Here we see that the velocity decreases due to an enhancement of the density.  The experimental data, for comparison, starts out with a center-of-mass velocity $\sim 1$ $\text{mm/s}$.  The impulse formula for Hubbard-band quasiparticles therefore  gives the correct order of magnitude and shows suppression due to a modulation of the density due to disorder.

\section{Temperature Dependence}
\label{sec_temp}

\begin{figure}[t]
\includegraphics[clip,width=80mm]{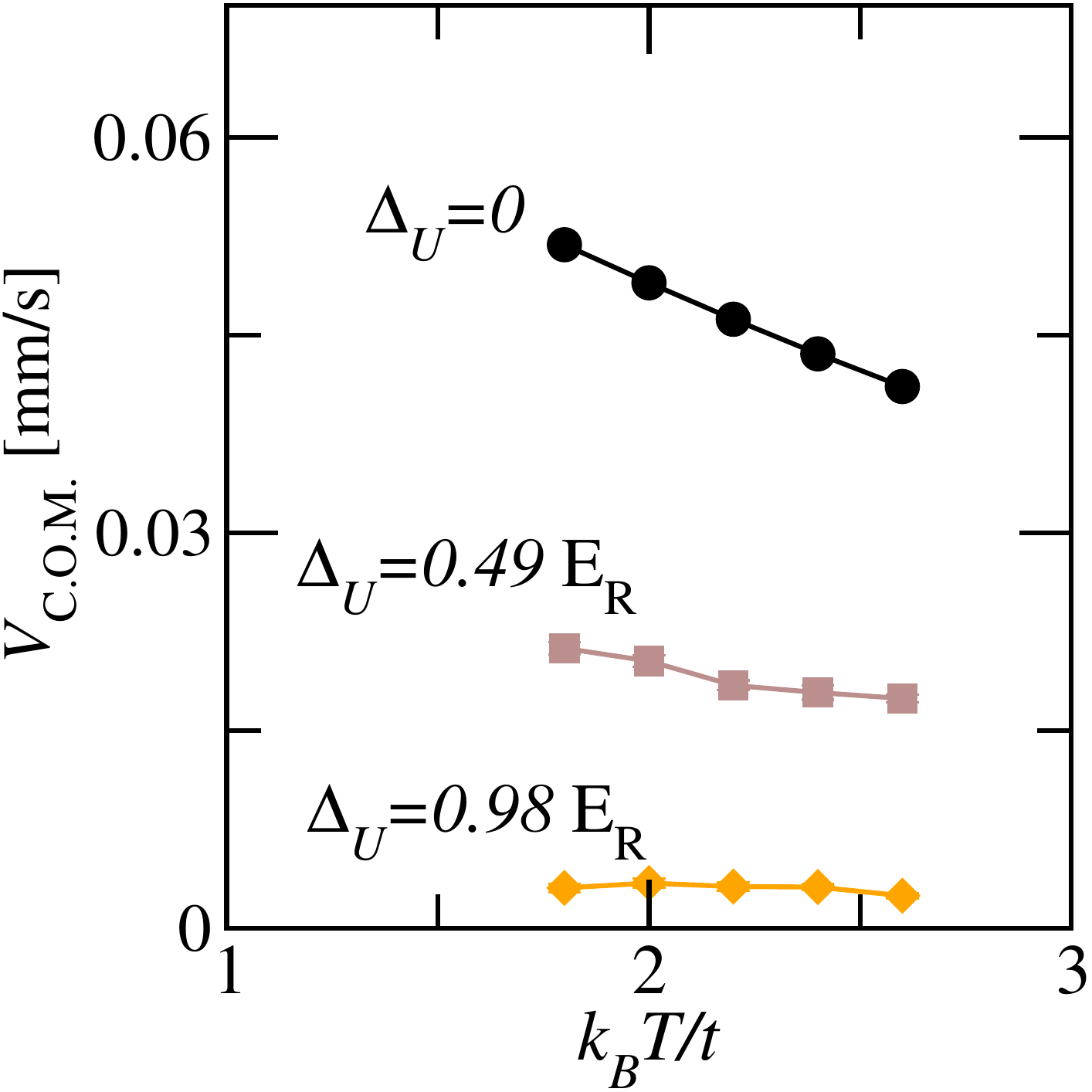}
\caption{(Color online) The disorder averaged center-of-mass velocity (Eq.~(\ref{eq_vcom_scalar})) as a function of temperature for several disorder strengths computed from solutions of  Eqs.~(\ref{eq_eomrho}) and (\ref{eq_eomGamma}).  Parameters are chosen to yield a small system size replica of the larger system implied by the parameters for the $7 E_{R}$ data in Table~\ref{tab1} (see text).  The velocities are disorder averaged using uniform disorder.    Here we see that increasing temperature suppresses the velocity only at low disorder strengths.  
}
\label{fig_v_vs_T}
\end{figure}

In this section we study the temperature dependence of the center-of-mass velocity in small trapped systems by solving for the dynamics of correlators using Eqs.~(\ref{eq_eomrho}) and (\ref{eq_eomGamma}).  Here it is shown that temperature increases (expected in adiabatic heating) suppress the center-of-mass velocity but only for low disorder strengths.

We can use Eqs.~(\ref{eq_eomrho}) and (\ref{eq_eomGamma}) to compute the center-of-mass dynamics in trapped systems on small system sizes.  We solve Eqs.~(\ref{eq_eomrho}) and (\ref{eq_eomGamma}) numerically.  The initial state is determined using Eqs.~(\ref{eq_initrho}) and (\ref{eq_initgamma}) within the local density approximation at fixed temperature. Fig.~\ref{fig_v_vs_T} shows example results for the center-of-mass velocity.  The simulations are carried out on a periodic cubic lattice with edges of size  $L=11$ where the trap zeroes the density at the the edges.  We consider a small system size replica of larger experimental parameters by choosing a stronger trap frequency $\hbar\omega/t=0.757$ but at the same chemical potential as that found for experimental system sizes, $\mu_{0}/t=3.8$.   $\tau_{P}=0.514 h/t$ is chosen by a trap-dependent rescaling discussed in Sec.~\ref{sec_com_dynamics}.  The entropy is allowed to vary but otherwise the remaining parameters are chosen from the $7 E_{R}$ data in Table~\ref{tab1}.

Fig.~\ref{fig_v_vs_T} shows that by increasing temperature, the center-of-mass velocity can decrease at low disorder.  This is the opposite of what is expected from variable range hopping in common regimes, e.g., in semiconductors, where the presence of a bath typically increases conductivity with increasing temperature.  Here we do not have an external bath.  At low disorder, increasing temperature suppresses the amplitude for particles to tunnel between neighboring sites, e.g., $t \langle \rho^{\sigma}_{l,l+1} \rangle$, in the initial state.  As a result, the center-of-mass velocity (which scales linearly with the nearest neighbor elements of the single-particle density matrix) is suppressed with increasing temperature.  The high disorder limit has a different behavior.  Here the dynamics is strongly suppressed by disorder and the thermal suppression of tunneling has little effect.  These qualitative trends show that, when we study the experimentally relevant fixed entropy case, adiabatic heating due to disorder will tend to suppress the center-of-mass velocity only at low disorder strengths.

\bibliography{paper.bib}

\end{document}